%% file: paper.tex
\documentclass[aps,prd,twocolumn,superscriptaddress,preprintnumbers,floatfix,nofootinbib,notitlepage,showkeys,showpacs]{revtex4-1}
\usepackage{multirow}
\usepackage{siunitx}

\usepackage{graphicx,times}
\usepackage{latexsym}
\usepackage{mathtools}

\usepackage{amsmath,amssymb,amsbsy,amsfonts}
\usepackage{array}
\usepackage{bm}
\usepackage{graphics}
\usepackage{mathrsfs}
\usepackage{xcolor}
\usepackage{cancel}
\usepackage[normalem]{ulem}

\usepackage{hyperref}
\usepackage[capitalise]{cleveref}

\usepackage{xcolor}
\usepackage{makecell}
\newcommand\TopRule{\Xhline{0.08em}}
\newcommand\HeaderRule{\Xhline{0.05em}}
\newcommand\MidRule{\Xhline{0.03em}}
\newcommand\BotRule{\Xhline{0.08em}}

\newcommand\SetTableProperties{
  \renewcommand{\arraystretch}{1.4}
}

\newcommand{\bfr}{{\mathbf{r}}}

\newcommand{\srr[1]}{{|s,\mathbf{r}#1\rangle}}
\newcommand{\mrr[1]}{{|m,\mathbf{r}#1\rangle}}
\newcommand{\srl[1]}{{\langle s,\mathbf{r}#1|}}
\newcommand{\mrl[1]}{{\langle m,\mathbf{r}#1|}}

\renewcommand\>{\rangle}
\newcommand\<{\langle}

\DeclareMathOperator\Tr{Tr}

\begin{document}

\title{\texorpdfstring{A qubit regularization of the $O(3)$ sigma model}{A qubit regularization of the O(3) sigma model}}
\author{Hersh Singh and Shailesh Chandrasekharan}
\affiliation{Department of Physics, Box 90305, Duke University, Durham, NC 27708, USA}

\begin{abstract}
We construct a qubit regularization of the $O(3)$ non-linear sigma model in two and three spatial dimensions using a quantum Hamiltonian with two qubits per lattice site. Using a worldline formulation and worm algorithms, we show that in two spatial dimensions our model has a quantum critical point where the well-known scale-invariant physics of the three-dimensional Wilson-Fisher fixed point is reproduced. In three spatial dimensions, we recover mean-field critical exponents at a similar quantum critical point. These results show that our qubit Hamiltonian is in the same universality class as the traditional classical lattice model close to the critical points. Simple modifications to our model also allow us to study the physics of traditional lattice models with $O(2)$ and $Z_2$ symmetries close to the corresponding critical points.\end{abstract}


\maketitle

\section{Introduction}
\label{sec1}

Quantitative understanding of quantum field theories pose unique computational challenges that we must overcome to be able to truly understand nature at a fundamental level \cite{Troyer:2004ge,Susskind:2014rva}. Currently, our understanding of these quantum many body theories is mainly obtained from perturbation theory. In a few cases when sign problems can be solved, quantum Monte Carlo methods can be used to compute equilibrium thermal averages and static ground state properties. However, in the vast number of cases involving non-equilibrium processes and in particular when theories are strongly coupled the available computational approaches are severely limited. One particularly promising approach to overcome this computational bottleneck is quantum computation \cite{Nielsen2011}. Universal quantum computers with tens of qubits already exist and it is likely that more advanced ones will begin to appear over the next decade. Anticipating this possibility, the field has exploded in recent years with new ideas and algorithms for using quantum computers to understand quantum many body systems and quantum field theories \cite{Jordan:2011ne,Jordan:2011ci,Casanova:2011wh,Casanova:2012zz,PhysRevX.6.031007,Kandala2017,Macridin:2018oli,Roggero:2018hrn}. Simple one dimensional quantum field theories are currently being studied extensively \cite{Zohar:2015hwa,Pichler:2015yqa,Martinez:2016yna,Banuls:2016gid,Klco:2018kyo,Kaplan:2018vnj,Frank:2019jzv}.

To study a quantum field theory using a quantum computer, in addition to the lattice regularization we need to formulate the theory with a local finite dimensional Hilbert space that can be represented with qubits. We will refer to this as the {\em qubit regularization} of a quantum field theory. In the traditional lattice regularization, local bosonic field operators satisfy the canonical commutation relations of the form $[\phi(x),\pi(y)] = i\delta_{x,y}$, which can only be realized with infinite dimensional representations. For this reason traditional scalar and gauge field theories are naturally formulated with lattice models that have infinite dimensional Hilbert space at every lattice site. Even the anti-commutation relations $\{\psi(x),\psi^\dagger(y)\} = \delta_{x,y}$ of fermionic field operators require special ideas to formulate using qubits \cite{Bravyi2002210,PhysRevA.95.032332}. We can define the qubit regularization of a quantum field theory as the construction of a quantum lattice Hamiltonian operator that acts on a finite dimensional Hilbert space at every spatial lattice site but reproduces the same physics as the traditional lattice regularized quantum field theory in the continuum limit. Like with traditional lattice regularization, continuum limits with qubit regularization also arise in the vicinity of quantum critical points. However, these critical points can be located in a region of parameter space which is not easily accessible in perturbation theory. In particular even Gaussian fixed points can require non-perturbative calculations. For this reason qubit regularizations have remained largely unexplored for many field theories.

An example of qubit regularization of quantum field theories is the D-theory approach \cite{Wiese:2006kp}.
In this approach we construct a $d+1$ dimensional quantum lattice Hamiltonian with a finite dimensional Hilbert space per lattice site. The extra dimension has a finite extent and is used as a coupling constant for the lower dimensional theory. At length scales much larger than the extent of the extra dimension, one obtains an effective $d$ dimensional quantum Hamiltonian (due to dimensional reduction) which acts as the qubit regularization of the original quantum field theory. Thus, the claim of the D-theory approach is that a $d+1$ dimensional quantum lattice Hamiltonian can provide the qubit regularization of a traditional $d+1$ dimensional lattice regularized quantum field theory. The extent in the extra dimension helps in creating the necessary Hilbert space of the problem at each $d$ dimensional spatial lattice site, while at the same time plays the role of a coupling of the theory. Asymptotically free two dimensional $CP(N-1)$ models have been formulated using this procedure more than a decade ago  Ref.~\cite{Beard:2004jr} and proposals for formulating lattice gauge theories and QCD appear even earlier \cite{Chandrasekharan:1996ih,Brower:2003vy}.

The goal of this work is to provide another concrete example of a qubit regularization in the context of the $O(3)$ sigma model in two and three spatial dimensions. Using this example we wish to show that sometimes a qubit regularization of a traditional $d+1$ dimensional lattice regularized quantum field theory is possible by constructing a $d$ dimensional lattice quantum Hamiltonian with a finite number of qubits per lattice site. In other words the dimensional reduction of the D-theory approach may not be necessary. By simply preserving the important symmetries the relevant continuum quantum field theory may emerge at an appropriate quantum critical point due to Wilson's renormalization group ideas. This approach is often used in condensed matter physics to argue that a particular $d+1$ dimensional quantum field theory naturally describes the long distance properties of a $d$ dimensional material. It was also recently advocated in \cite{Alexandru:2019ozf}. In our work, using quantum Monte Carlo methods, we show explicitly that our qubit Hamiltonian reproduces the critical scaling of the $O(3)$ Wilson Fisher fixed point and the Gaussian fixed point in two and three spatial dimensions respectively. We also show that simple modifications of our qubit Hamiltonian allows us to obtain the critical scaling of similar fixed points with $O(2)$ and $Z_2$ symmetries.

Another outcome of qubit regularizations is that they can lead to new ways of formulating Euclidean lattice field theories, especially within the worldline approach \cite{Chandrasekharan:2008gp}. In our work we show that the physics of the qubit regularized $O(3)$ model can be viewed from the perspective of the Hamiltonian formulation in continuous time or a relativistic lattice formulation in discrete time. Both view points reproduce the expected scaling at the critical points. However, the relativistic limit leads to a much simpler worldline approach than the traditional lattice regularized models. In other words our relativistic models are simplified versions of the dual formulations of $O(N)$ models constructed recently \cite{Wolff:2009kp,Bruckmann:2015sua,Bruckmann:2016txt,Gattringer:2017hhn}. A similar simplified relativistic model with $O(4)$ symmetry was studied in \cite{Banerjee:2019jpw}. Such models have also been constructed to study condensed matter phenomena \cite{kaul-spin_nimetics-2015}. 

This paper is organized as follows. In \cref{sec:qubit-model}, we construct our qubit Hamiltonian for the $O(3)$ model and show how to construct its worldline formulation using path integrals in \cref{sec:worldline-formulation}. We distinguish between the Hamiltonian limit and the relativistic limit. In \cref{sec:worm-algorithm}, we briefly sketch the worm algorithm and discuss the observables we measure. Then in \cref{sec:results} we present our results and discuss our conclusions in \cref{sec:conclusions}.

\begin{figure}[t]
    \centering
    \includegraphics[width=\linewidth]{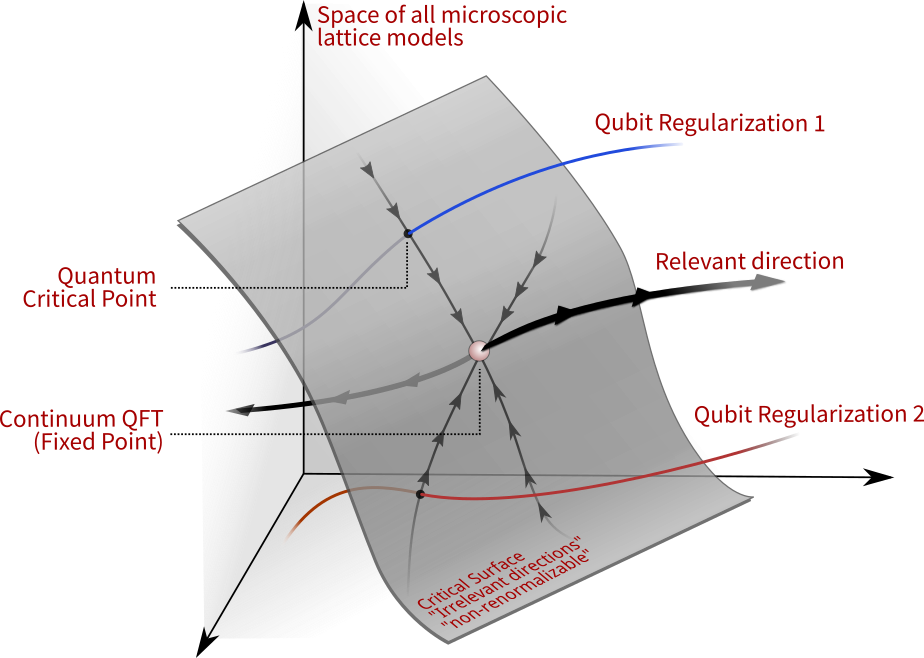}
    \caption{Schematic of how qubit regularizations fits into the usual picture of Wilson's renormalization group ideas. The two lines shown as Qubit Regulatization 1 and Qubit Regularization 2 show a set of qubit Hamiltonians where one parameter is varied. These are not RG flow lines, which are shown with arrows.}
    \label{fig:qubit-regularization}
\end{figure}

\section{The Qubit Model}
\label{sec:qubit-model}

Our goal is to construct the qubit regularization for the continuum quantum field theories that emerge from the traditional lattice regularized classical non-linear $O(3)$ sigma model, whose action is given by
\begin{align}
S = -\frac{1}{g} \sum_{\langle ij \rangle} \vec{\phi}_i \cdot \vec{\phi}_{j}, 
\label{eq:tradmodel}
\end{align}
where $i$ and $j$ are nearest neighbor sites on a Euclidean lattice site in $d+1$ space-time dimensions and $\vec{\phi}_i$ is a classical unit $3$-vector associated to that site. Continuum limits of lattice field theories emerge at second order critical points $g_c$ of the lattice model. In this case, \cref{eq:tradmodel} has one such critical point  separating the broken phase from a symmetric phase in both $d=2$ and $d=3$. In $d=2$, we obtain a conformal field theory governed by the $O(3)$ Wilson-Fisher fixed point and in $d=3$, we obtain the physics of the Gaussian fixed point (free field theory). In this work, we reproduce the physics close to these two fixed points using a $d$-dimensional quantum Hamiltonian
with two qubits per lattice site.  We first construct a quantum Hamiltonian with a global $O(3)$ symmetry and later extend our model to an $O(2)$ or a $Z_2$ model.

Our model is defined on a regular $d$ dimensional periodic spatial lattice with $L$ sites in each direction.  At each spatial site $\bf r$, we have a singlet state $\srr[]$ and three triplet states $\mrr[]$ ($m=0,\pm1$), which form the four orthonormal basis states of two qubits.  The singlet acts as the Fock vacuum while the triplets carry the $O(3)$ charge. The Hamiltonian of our model is defined as a sum of two terms,
\begin{align}
 H & = H_1 + H_2
\label{eq:latticemodel}
\end{align} 
where $H_1$ is a sum over single-site operators and $H_2$ is a sum over nearest-neighbor operators. The first term is given by
\begin{align}
H_1 &= \sum_\bfr \Big( J_t H^t_\bfr - \mu Q_\bfr\Big),
\label{eq:siteH}
\end{align}
where $J_t$ is a coupling, $H^t_\bfr$ is a projection operator onto the triplet $|m\>$ states on the site $\bfr$ ,
\begin{align}
H^t_\bfr \ &=\ \sum_m \mrr[]\mrl[]
\label{eq:bondH}
\end{align}
and $Q_\bfr$ is the $O(3)$-charge operator given by
\begin{align}
Q_\bfr \ &=\ \sum_m \ m\ \mrr[]\mrl[],
\end{align}
with $\mu$ the chemical potential.
For $\mu > 0$, the $|m=1, \mathbf{r} \>$ states are enhanced and the $|m=-1, \mathbf{r}\>$ states are suppressed.  The second term in the Hamiltonian is 
\begin{align}
H_2 &= -\sum_{\langle \bfr,\bfr'\rangle}\ \Big( J_h H^h_{\bfr,\bfr'} + J_p H^p_{\bfr,\bfr'} \Big),
\end{align}
where $J_h$ and $J_p$ are couplings, and $H^h_{\bf r,\bf{r'}}$ and $H^p_{\bfr,\bf{r'}}$ are bond operators on the link connecting the nearest neighbor sites $\bf r$ and $\bf{r'}$.  These bond operators act on a $16$-dimensional state space with the basis vectors $\{\srr[]\srr[']$, $\srr[]\mrr[']$,  $\mrr[]\srr['], 
\mrr[] |m'\bfr'\> \}$. The term $H^h_{\bf r,\bfr'}$ is the hopping part of the Hamiltonian and is given by
\begin{align}
H^h_{\bfr,\bfr'} = \sum_m  \Big\{
&\srr[]\mrr['] \mrl[]\srl['] \nonumber\\
&+ \mrr[]\srr['] \srl[]\mrl['] \Big\} 
\end{align}
while $H^p_{\bfr,\bfr'}$ denotes the pair creation/annihilation events and takes the form
\begin{align}
H^p_{\bfr,\bfr'} = \sum_m \Big\{ 
& \mrr[] |{-m},\bfr'\rangle \srl[]\srl[']  \nonumber\\
& + \srr[]\srr['] \mrl[] \< {-m},\bfr'| \Big\}.
\end{align}
For convenience, we choose $J_h=J_p=J$ in this work, although this restriction is not necessary to preserve the symmetries of interest.

\begin{figure}[t]
    \centering
    \includegraphics[]{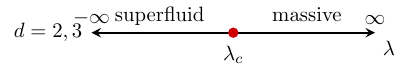}
    \caption{The zero temperature phase diagram of our qubit Hamiltonian in $d=2,3$.}
    \label{fig:phasediag}
\end{figure}
    
When $\mu=0$, our model \cref{eq:latticemodel} has a global $SU(2)$ symmetry under which all qubits in the model transform as a spin-half state.  
Under these transformations, $\srr[]$ is invariant by definition and the triplet states $\mrr[]$ ($m = 0,\pm 1$) transform as the spin-$1$ representation of $SU(2)$,
\begin{align}
\mrr[] \rightarrow \sum_{m'} D^{(1)}_{m,m'} |m',\bfr\rangle,
\end{align}
where $D^{(1)}_{m,m'}$ are the $SO(3)$ rotation matrices in the basis which diagonalizes the $z$ the generator of rotations around the $z$ axis.  
This makes all three terms $H^t_\bfr$, $H^h_{\bfr,i}$ and $H^p_{\bfr,i}$ invariant. 
The chemical potential $\mu$ breaks the $O(3)$ symmetry, but we can use it to measure the mass of the $O(3)$ particles if needed. In this work, we set $\mu=0$.

It is straightforward to extend the our model to obtain qubit regularizations of $O(2)$ and $Z_2$ quantum field theories in two and three dimensions. Consider, for example, adding the on-site term
\begin{align}
H_3 = J_z \sum_{\bf r} \Big\{ 
|0,\mathbf{r}\> \<0,\mathbf{r}| - \sum_{\mathclap{m=\pm 1}} |m,\mathbf{r}\> \<m,\mathbf{r}| \Big\}
\end{align}
to the Hamiltonian in \cref{eq:latticemodel}. When $J_z\rightarrow \infty$, the $m=0$ states are forbidden from the theory and the qubit model is only invariant under the $SO(2)$ subgroup,
\begin{align}
    |m, {\bf r}\> \to e^{i\theta m} |m, {\bf r}\>.
\end{align}
and $Z_2$ transformations $|m, {\bf r}\> \to |-m, {\bf r}\>$. Thus the symmetry group of our model is reduced to $O(2)$, and it should naturally provide a qubit regularization of the $O(2)$ sigma model. On the other hand, when $J_z\rightarrow -\infty$, the $|m=\pm 1\>$ states are forbidden and only the $|m=0\>$ states are allowed. In this case, we get a model which is invariant under the $Z_2$ transformation $|m=0, {\bf r}\> \to - |m=0,{\bf r}\>$. We then obtain a qubit regularization of the real scalar field theory.

The phase structure of our qubit regularized model in \cref{eq:latticemodel} can easily be understood in terms of the dimensionless coupling $\lambda = J_t/J$ (see \cref{fig:phasediag}). When $\lambda \rightarrow \infty$, the lattice Hamiltonian is in a symmetric massive phase since the $\mrr[]$ states are suppressed and the singlet states $\srr[]$ dominate. On the other hand when $\lambda \rightarrow -\infty$, every space-time lattice site contains a triplet state $\mrr[]$, and this most likely leads to spontaneous symmetry breaking. Assuming there is a second order quantum critical point at some intermediate coupling $\lambda_c$, according to Wilson's renormalization group ideas, the continuum quantum field theory that emerges close to $\lambda_c$ on either side would be the same as that of the traditional lattice model \eqref{eq:tradmodel}. For example, if we focus on the theory for $\lambda > \lambda_c$, we obtain the symmetric massive phase of the $O(3)$ sigma model where we can use the mass scale to set the lattice spacing. In $3+1$ dimensions, this theory will be free up to logarithmic corrections, while in $2+1$ dimensions, the scaling of the theory will be described by the Wilson-Fisher fixed point. These arguments extend to $O(2)$ and $Z_2$ cases as well. In this work, we show explicitly, using Monte Carlo calculations, that the scaling properties of the traditional model are reproduced by our qubit regularized model.

\begin{figure*}[htb]
\centering
\includegraphics{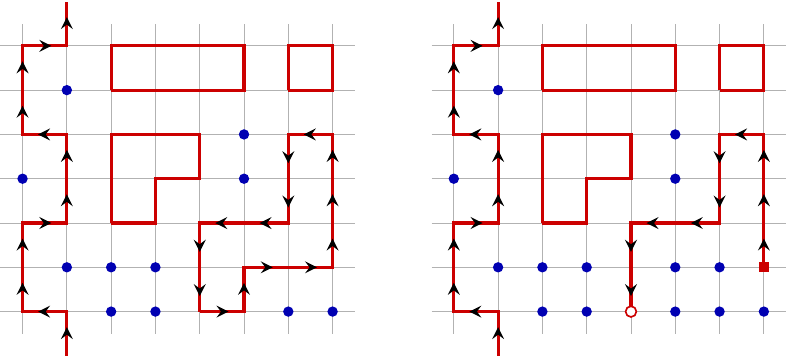}
\caption{Illustration of a worldline configuration (left) and a worm (or defect) configuration (right) in one spatial dimension. The worm is shown as the thick solid line with a head (open circle) and a tail (filled square). The lines without a head or a tail show particle worldlines, while sites with a Fock vacuum (singlets) are shown as filled circles. Particle with charge $m=\pm1$ are shown as oriented worldlines while those with charge $m=0$ are unoriented. }
\label{fig:worldline}
\end{figure*}

\section{Worldline formulation}
\label{sec:worldline-formulation}

In order to show that our qubit regularization reproduces the same physics as the traditional lattice regularization near the critical point, we compute observables within our qubit model using Monte Carlo methods and study their scaling properties near the critical point. We do this using the worldline approach and ideas of worm algorithms for updating the configurations \cite{Chandrasekharan:2006tz,Cecile:2007dv,PhysRevD.99.074511}. The partition function of our model $Z=\mathrm{Tr}(\mathrm{e}^{-\beta H})$ can be expanded as 
\begin{align}
Z = \sum_k \int [dt_k... & dt_1] \Tr
\Big(e^{-(\beta-t_k) H_1} (-H_2) \nonumber \\
&  e^{-(t_k-t_{k-1}) H_1} \cdots (-H_2) e^{-(t_1) H_1} \Big),
\end{align}
where we treat $H_1$ as a free term and $H_2$ as a perturbation. However, the integer $k$, which labels the number of insertions of $H_2$ terms, is allowed to take any value and hence the above expansion is not an approximation.  Inserting the expression for $H_2$ as a sum over nearest neighbor bond operators $H_b^\sigma$ (either $J_h H_{\bfr,i}^h$ or $J_p H_{\bfr,i}^p$), we can rewrite the above expression as sum over $k$ bond configurations $[b,\sigma]$ at times $t_1,..,t_k$ given by 
\begin{align}
Z = \sum_k \int [dt_k... & dt_1] \sum_{[b,\sigma]} \Tr
\Big(e^{-(\beta-t_k) H_1} (-H_{b_k}^{\sigma_k}) \nonumber \\
&  e^{-(t_k-t_{k-1}) H_1} \cdots (-H_{b_1}^{\sigma_1}) e^{-(t_1) H_1} \Big).
\end{align}
We can evaluate the trace in the singlet-triplet basis by inserting a complete set of states after every insertion of the bond operator which can have non-zero off diagonal matrix elements. This then leads to a worldline configuration depicting the motion of $m = 0,\pm 1$ type particles in a Fock vacuum ($s$ type sites).

For convenience, we also discretize time into $L_T$ equal parts with temporal extent $\varepsilon$ (that is, $\varepsilon L_T = \beta$) and map the worldline configuration onto a space-time lattice. The partition function then takes the form
\newcommand\Conf{n}
\begin{align}
Z \ =\ \sum_{[\Conf(\vec r, t)]} \prod_{\langle ij\rangle} \ W_{\langle ij\rangle}
\label{eq:latpf}
\end{align}
where the sum is over all lattice worldline configurations
$[\Conf(\vec r, t)]$ with $\Conf(\vec r, t)=\{s,m=0,\pm 1\}$ at each lattice site $(\vec r, t)$ 
and $W_{\langle ij\rangle}$ are weights associated with all space-time bonds $\langle ij\rangle$. The configuration $[\Conf(\vec r, t)]$ is composed of vacuum sites (no particles) and sites where one particle of type $m=0,\pm 1$ is moving. 
\Cref{fig:worldline} is an illustration of a worldline configuration in $1+1$ dimensions. Particle world lines are shown with lines on the bonds connecting lattice sites and vacuum sites are depicted as sites with filled circles. Each particle worldline is a loop, that may be oriented (depicting $m=\pm 1$ type particles) or unoriented (depicting $m=0$ type particles). A temporal bond that contains a $m=+1$ ($m=-1$) particle worldline moving through it is depicted by an arrow pointing in the positive (negative) time direction. The weights $W_{\langle ij\rangle}$ can be computed by looking at the configuration on the bond $\langle ij\rangle$. 
If the bond $\langle ij\rangle$ is empty, then $W_{\langle ij\rangle} = 1$, otherwise the weight depends on whether the bond is along a spatial direction or a temporal direction. For convenience, we define three weights 
\begin{align}
W_s = \varepsilon J,\quad W_t = e^{-\varepsilon J_t},\quad W_\mu =e^{\mu\varepsilon}. 
\label{eq:weights}
\end{align}
If the bond contains a particle worldline along the spatial direction then $W_{\langle ij\rangle} = W_s$, but if it is along the temporal direction then $W_{\langle ij\rangle} = W_t (W_\mu)^m$. The latter term also depends on the $O(3)$ charge of the particle on the temporal bond.

In the qubit regularization that we are propose here, rotational symmetry between space and time is not always guaranteed. Thus, it is usually difficult to understand how one can recover a continuum relativistic quantum field theory in this approach.  Here we rely on the fact that close to quantum critical points the long distance theory may flow to a quantum field theory that is naturally relativistically invariant. Fortunately, in our worldline formulation it is easy to see that relativistic invariance is indeed recovered.  If we set $W_s = W_t$, our worldline formulation becomes invariant under space-time lattice rotations. Thus, by setting $\varepsilon=1$ and $\exp(-\varepsilon J_t) = \varepsilon J$, we are guaranteed that the quantum critical point obtained by tuning $J$ will be relativistically invariant. We will refer to this as the `relativistic limit' of our qubit regularized lattice field theory. This is in contrast to the `Hamiltonian limit' which is obtained in the time continuum limit $\varepsilon \to 0$ and by tuning $\lambda = J_t/J$ to locate the critical point.

From the perspective of implementing a qubit formulation of quantum field theories on a quantum computer, we are more interested in the Hamiltonian limit. As stated above, in this limit there is no symmetry between space and time and it becomes difficult to argue that we will recover relativistic invariance near the quantum critical point. However, given that our model has a quantum critical point in the relativistic limit, it very likely that this critical point survives in the Hamiltonian limit.  We can, in principle, formulate an algorithm directly in the time continuum limit ($\varepsilon \rightarrow 0$) and compute quantities as a function of  $\lambda$. However, in this work, we choose $\varepsilon = 0.1$ for convenience, and refer to the results as the Hamiltonian limit.

\begin{figure*}[h]
    \centering
    \includegraphics[]{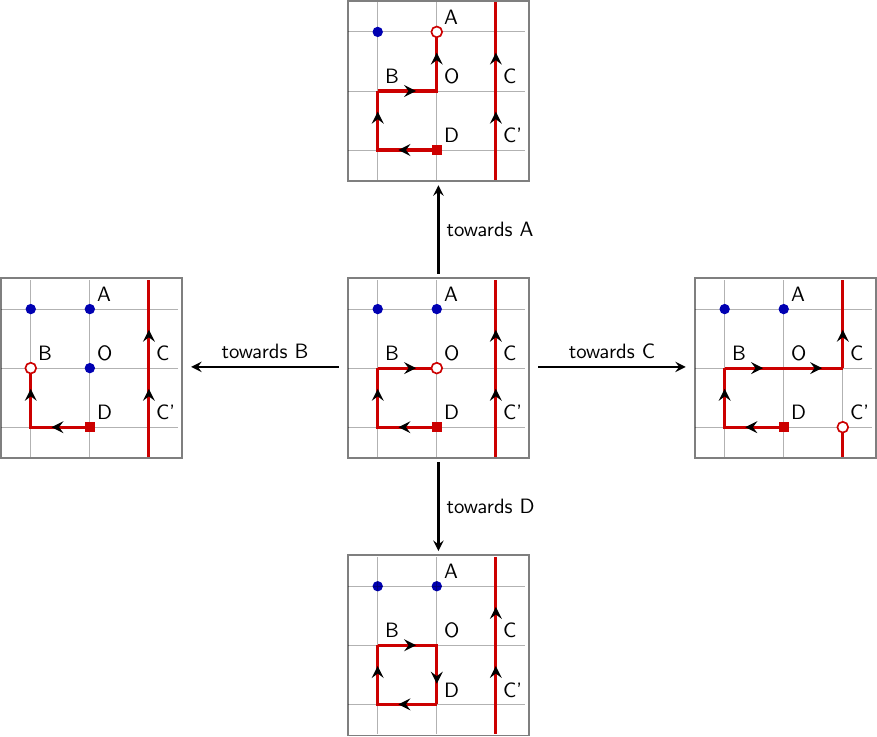}
    \caption{Possible local moves for the worm head. The configurations shown have the same convention as  \cref{fig:worldline}. Starting from the configuration shown in the center, the worm head can now move to one of the four positions $A,B,C,D$.  We choose the probability of each move to satisfy detailed balance.}
    \label{fig:worm-updates}
\end{figure*}

\section{Worm Algorithm}
\label{sec:worm-algorithm}

A worm algorithm to update our worldline configurations $[n({\bf r},t)]$, described in \cref{sec:worldline-formulation}, can easily be constructed by extending ideas developed previously for similar worldline models \cite{Chandrasekharan:2006tz,Cecile:2007dv,PhysRevD.99.074511}. The basic idea is to sample the worldline configurations along with configurations with two defects in the form of one creation operator $a^\dagger_{\bfr,m} = \mrr[]\srl[]$ and one annihilation operator $a_{\bfr,m} = \srr[]\mrl[]$. These configurations with defects  are referred to as worm configurations (illustrated in \cref{fig:worldline}). The location of the creation (annihilation) operator is regarded as the tail (head) of the worm. Although the defect configurations are different from the configurations that contribute to the partition function, the rules to compute their weights are the same. In particular, the defects do not carry any new weights.

Since the construction of worm algorithms is well established by now, we simply sketch the main ideas here.
The algorithm begins with creating a head and a tail on the same site or nearest neighbor sites and propagating the head locally around the lattice until it reaches the tail and can be removed. Each of the local moves satisfies detailed balance.  For efficiency, the local moves are designed not to retrace the steps backwards, if possible while maintaining detailed balance, so that the worm can explore new regions of the parameter space faster. \Cref{fig:worm-updates} shows a local worm configuration in the center and exhibits the four possible types of local moves for the worm head on a two-dimensional spacetime lattice. In general there will be $2(d+1)$ local moves that are proposed and accepted according to detailed balance. In the example shown, the worm head $O$ can move to one of four labelled positions $A,\dotsc, D$. We choose one of these four directions for the worm head to move randomly. The simplest case is towards site $A$:  we simply propose to create a bond $OA$ in that direction by removing the monomer on site $A$ and moving the worm head there. If the direction is towards site $B$, we propose to delete the bond $OB$, create a monomer on site $O$ and move the worm head back to position $B$.  On the other hand, if the worm head decides to move towards position $C$, we propose to create a bond $OC$, and delete the bond $CC'$, moving the worm head to position $C'$. This is a two site move for the worm head because we do not allow configurations with more than one two bonds at each site. Finally, the site $D$ contains the worm tail. 
So, if the direction chosen is towards site $D$, we propose to close the worm by creating the bond $OD$, thereby transforming the worm configuration transforms into a regular worldline configuration. If this proposal is accepted according to detailed balance the worm update ends. Note each of the four local steps can be retraced and hence the probabilities that satisfy detailed balance can be worked out,

In our qubit models we have both oriented and unoriented loops. Although the illustration above involved the worm head and the worm tail to be on an oriented loop, a similar strategy can be adapted for the unoriented loop. The move towards $C$ then leads to two possible moves for the head. We have constructed worm algorithms that update each of the sectors separately. In other words, while the oriented loops are being updated the unoriented loops are frozen and vice versa. We then also add a simple metropolis update which flips between the two types of loops. This loop-flip update then makes the entire algorithm ergodic. In the $O(3)$ case in fact we do not even perform the worm update on the unoriented loops, but combine the worm algorithm on the oriented loops with a loop-flip update. In the $O(2)$ ($Z_2$) case there are no unoriented (oriented) loops and hence there is no need to perform the loop-flip update.

Using the above worm algorithm we can compute three observables as discussed below.
\begin{enumerate}
\item The first observable is the average density of vacuum sites $v$ which we define as
\begin{align}
v &= \frac{1}{Z} \mathrm{Tr}\Big( \frac{1}{L^d} \sum_\bfr\ P^s_\bfr \mathrm{e}^{-\beta H} \Big)
\end{align}
Given a configuration $[n({\bf r},t)]$ this quantity can easily be computed by computing the sites with vacuum sites on them.
\item 
The second observable is the average $O(3)$ charge defined as
\begin{align}
\langle Q \rangle = \frac{1}{Z} \mathrm{Tr}\Big( \sum_\bfr\ Q_\bfr \mathrm{e}^{-\beta H} \Big).
\end{align}
In each worldline configuration the $O(3)$ charge $Q_\bfr$ is a conserved quantity and does not change in time and can be easily computed. When $\mu=0$, we expect $\langle Q\rangle=0$ due to the $SO(3)$ symmetry. However, as $\mu$ increases $\langle Q\rangle$ will increase and cross $0.5$ at a critical coupling $\mu_c$. In the massive phase when $\beta, L \rightarrow \infty$ this critical coupling gives the mass of the $O(3)$ particles. This technique to computing the mass has also been used with traditional formulations \cite{Bruckmann:2016txt}.

\item The second observable is the current-current susceptibility (which is related to the superfluid density) $\rho_s$ defined through the $O(3)$ conserved current. One can compute it using the conserved charge $Q_w$ along one of the spatial directions for every worldline configuration $[n({\bf r},t)]$, using the formula
\begin{align}
\rho_s \ =\ \frac{1}{L^{d-2} \beta} \langle Q_w^2 \rangle,
\end{align}
where the average is computed in the worldline formulation.
\item Our final observable is the susceptibility of the two point correlation function involving the creation and annihilation of particles. This is given by
\begin{align}
\chi & = \frac{1}{Z L^d} \sum_{\bfr,\bfr'} \int_0^\beta dt \ 
\mathrm{Tr}\Big( \mathrm{e}^{-(\beta - t) H} a_{\bfr,m} \mathrm{e}^{-t H} a^\dagger_{\bfr',m}\Big). \nonumber \\
\label{eq:chidef}
\end{align}
Computing $\chi$ is straight forward in our worm algorithm since the during the worm update we naturally sample configurations with a creation and annihilation event (see \cref{fig:worldline}). 
\end{enumerate}

In the next section we discuss our results for the three observables close to the critical point separating the symmetric phase from the broken phase in both $d=2$ and $d=3$, for the $O(3)$, $O(2)$ and $Z_2$ cases.

\begin{figure*}[hptb]
\centering
\newcommand\WidthFactor{0.95}
\includegraphics[width=\WidthFactor\linewidth]{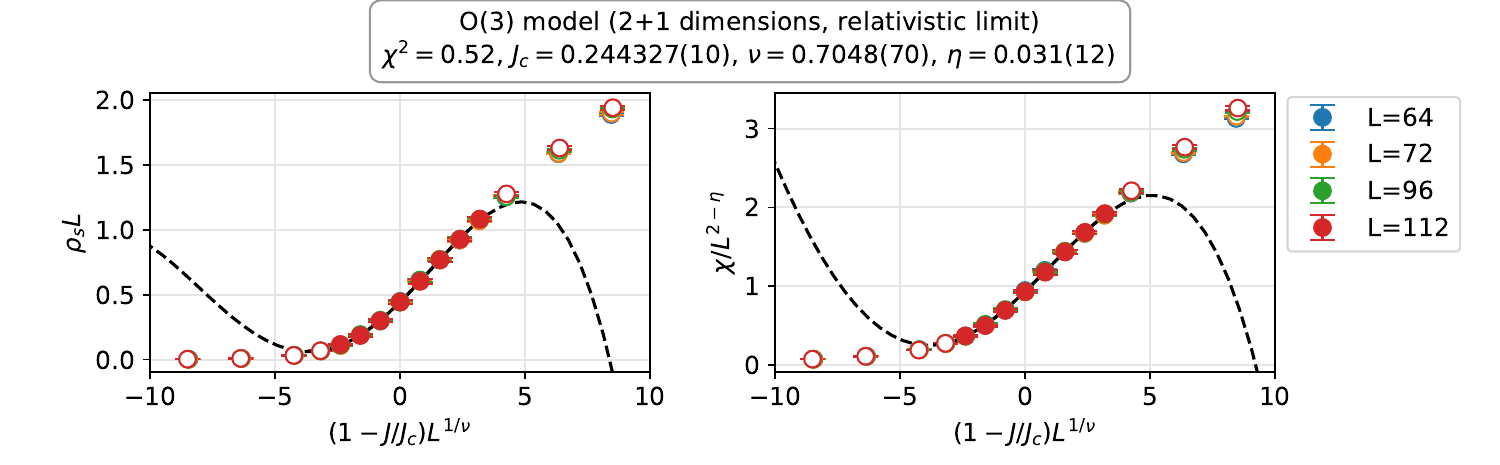}\\
\includegraphics[width=\WidthFactor\linewidth]{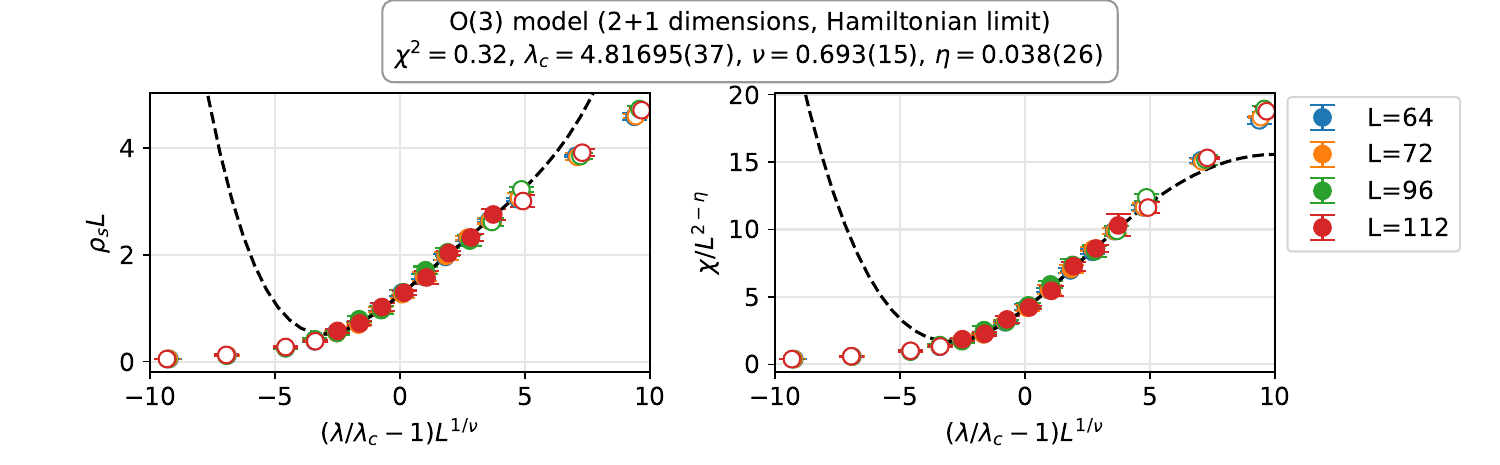}\\
\includegraphics[width=\WidthFactor\linewidth]{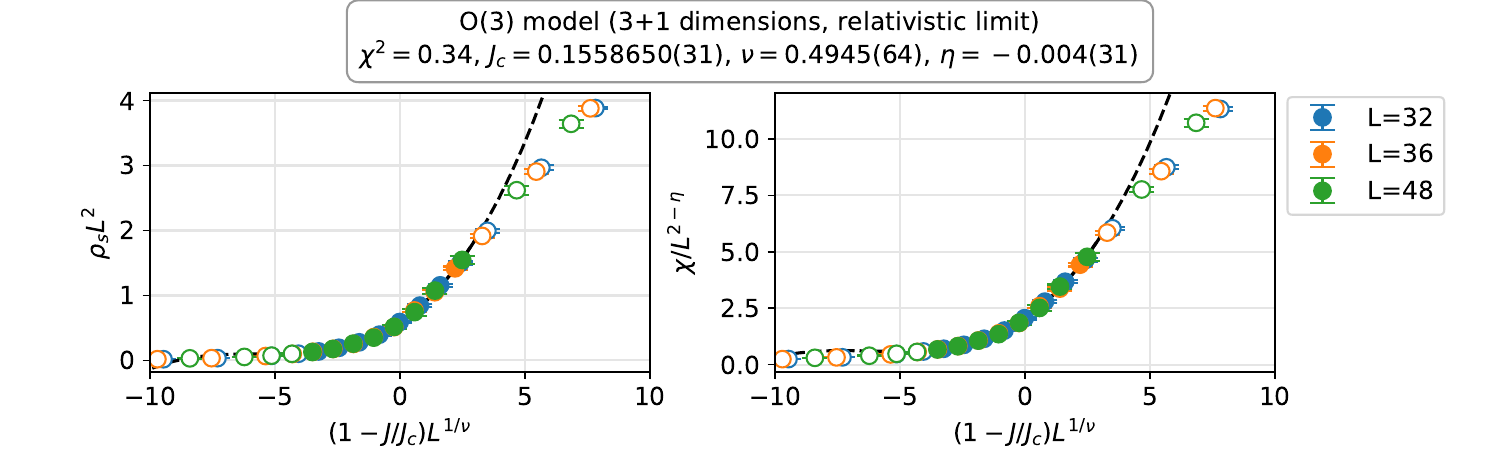}\\
\includegraphics[width=\WidthFactor\linewidth]{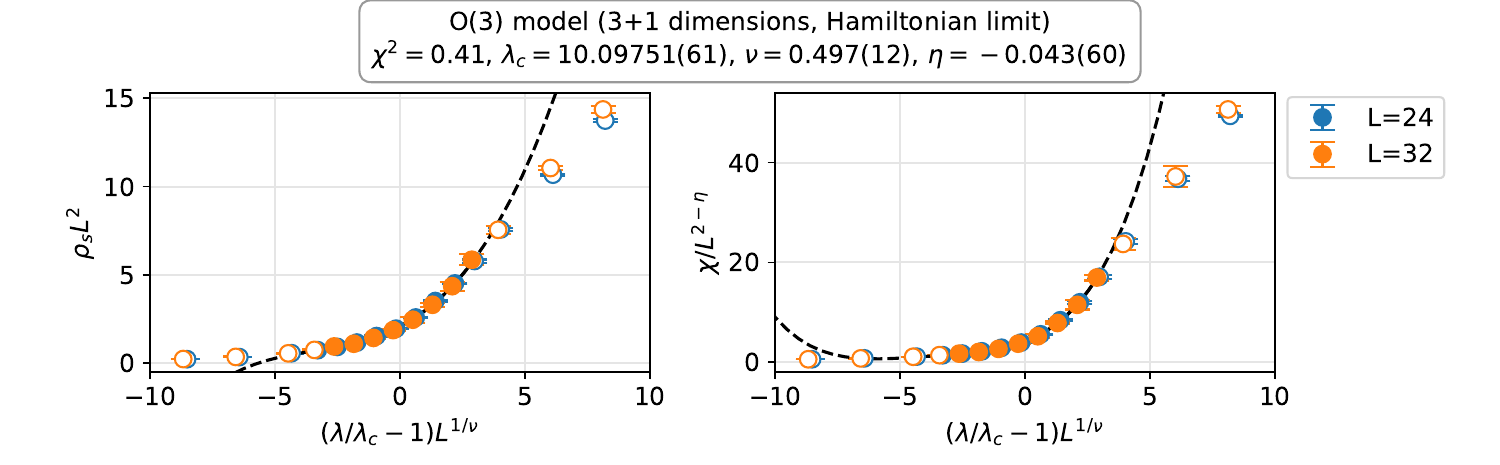}
\caption{Results for the $O(3)$ qubit model.  Plots of $\rho_s L^{d-1}$ and $\chi/L^{2-\eta}$ as a function of $(J-J_c) L^{1/\nu}$ (in the relativistic limit) and as a function of $(\lambda-\lambda_c) L^{1/\nu}$ (for the Hamiltonian limit).  The black line shows a combined fit in each case assuming that $f(x)$ and $g(x)$ in \cref{eq:critical-scaling} can be approximated by a polynomial up to fourth order.
}
\label{fig:O3-critical-scaling}
\end{figure*}

\begin{figure*}[hptb]
\centering
\newcommand\WidthFactor{0.95}
\includegraphics[width=\WidthFactor\linewidth]{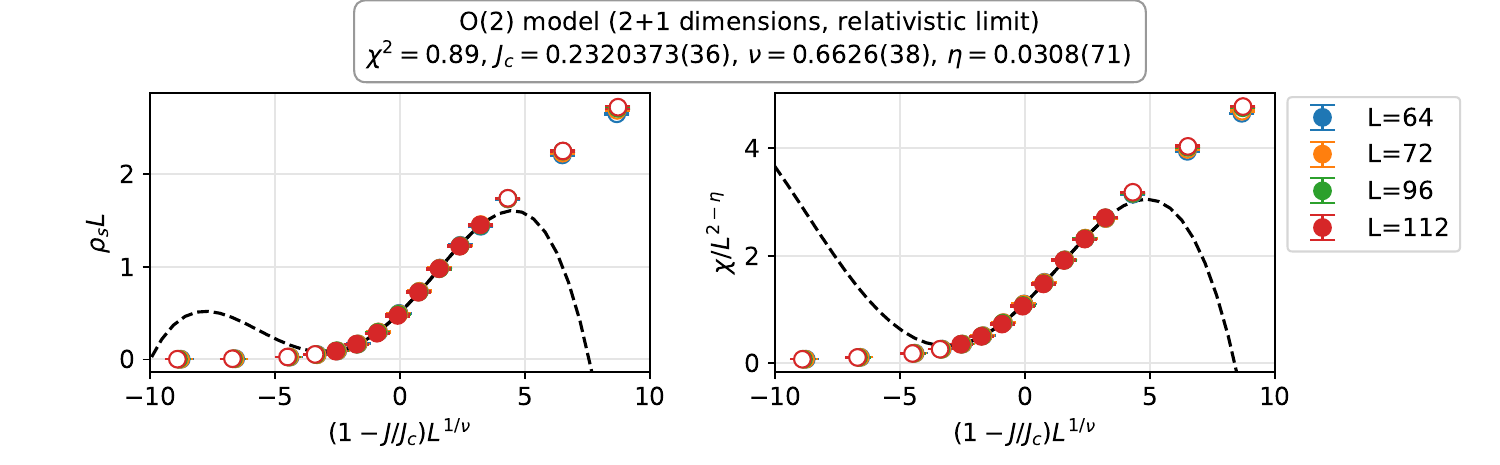}\\
\includegraphics[width=\WidthFactor\linewidth]{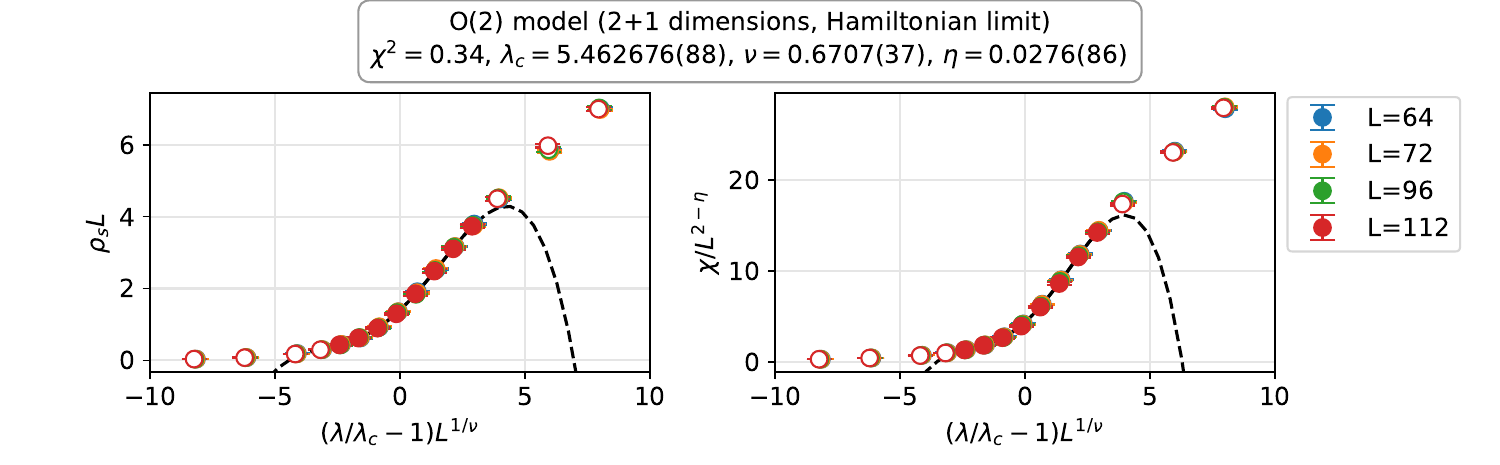}\\
\includegraphics[width=\WidthFactor\linewidth]{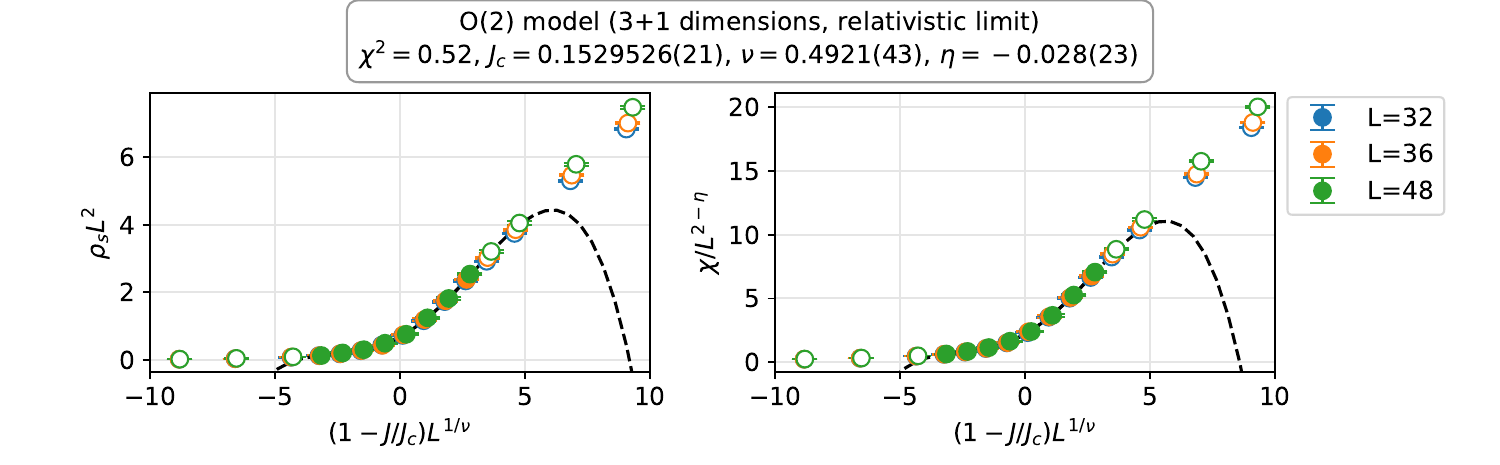}\\
\includegraphics[width=\WidthFactor\linewidth]{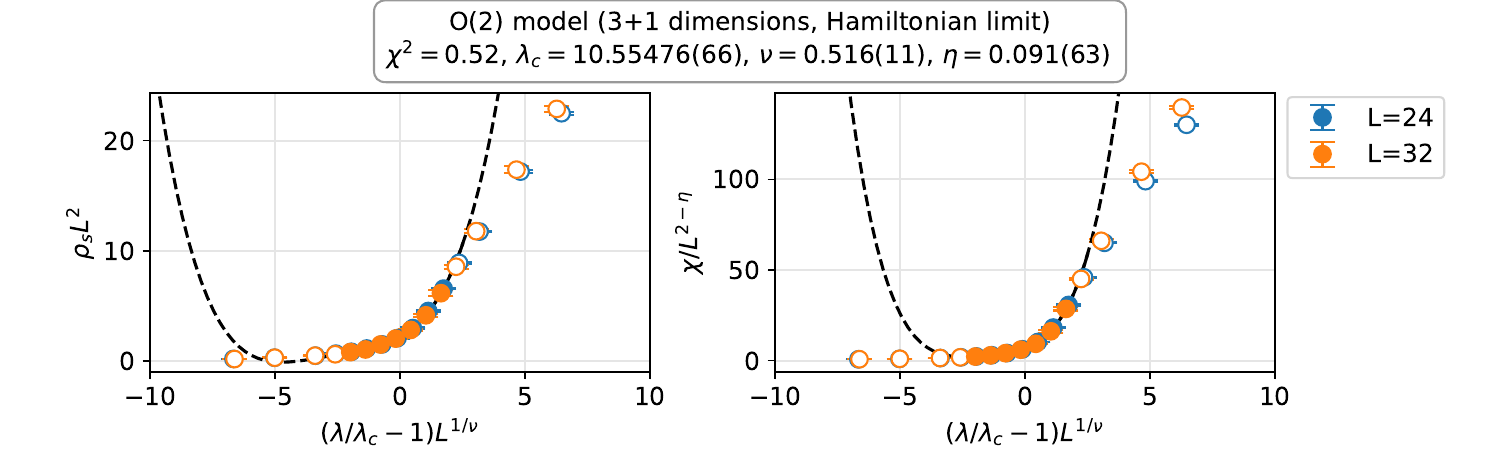}
\caption{Results for the $O(2)$ qubit model.  Plots of $\rho_s L^{d-1}$ and $\chi/L^{2-\eta}$ as a function of $(J-J_c) L^{1/\nu}$ (in the relativistic limit) and as a function of $(\lambda-\lambda_c) L^{1/\nu}$ (for the Hamiltonian limit) in $d=2,3$ dimensions.  The black line shows a combined fit in each case assuming that $f(x)$ and $g(x)$ in \cref{eq:critical-scaling} can be approximated by a polynomial up to fourth order.}
\label{fig:O2-critical-scaling}
\end{figure*}

\begin{figure*}[htb]
\centering
\newcommand\WidthFactor{0.49}
\includegraphics[width=\WidthFactor\linewidth]{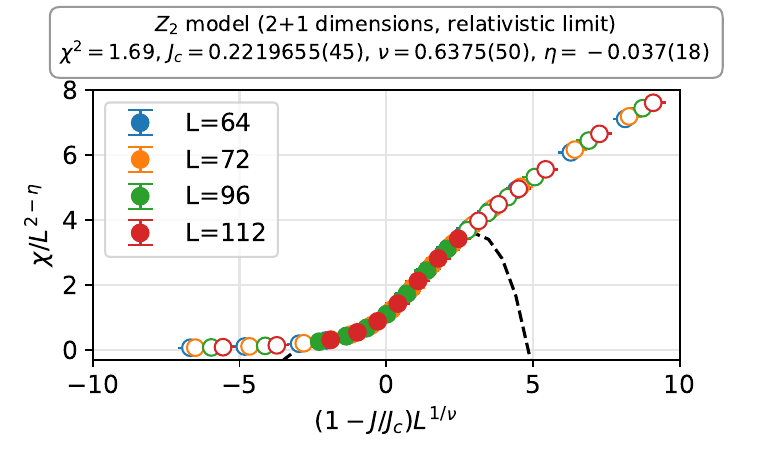}
\includegraphics[width=\WidthFactor\linewidth]{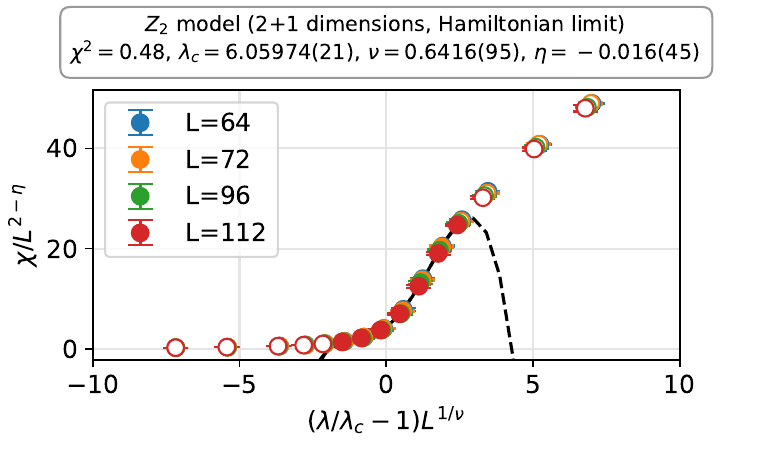}\\
\includegraphics[width=\WidthFactor\linewidth]{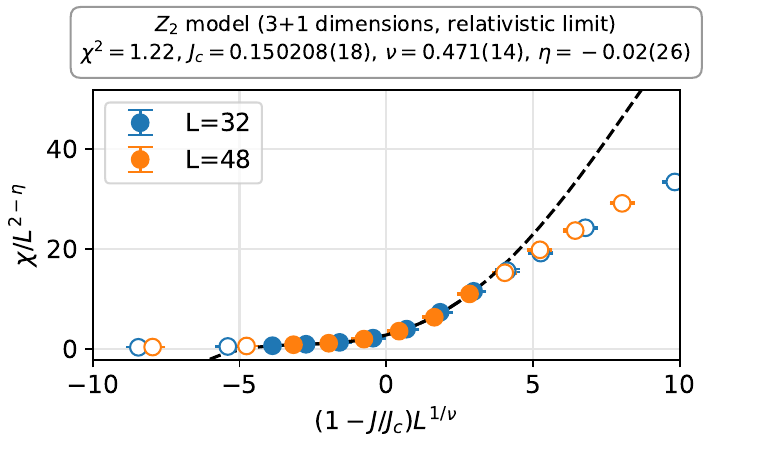}
\includegraphics[width=\WidthFactor\linewidth]{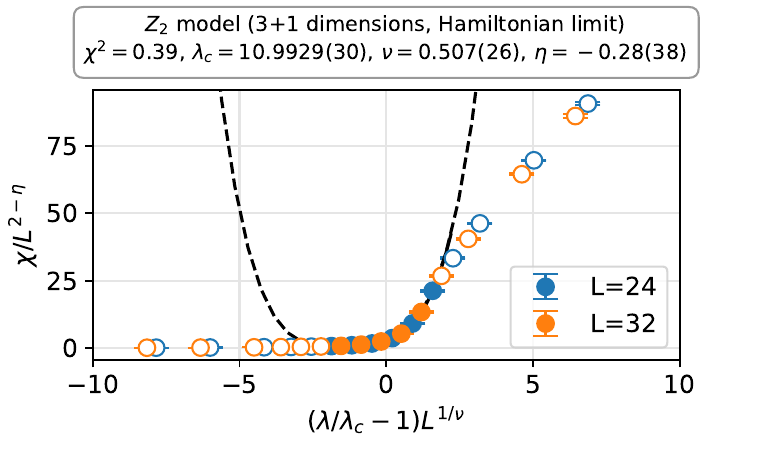}\\
\caption{Results for the $Z_2$ qubit model. Plots of $\chi/L^{2-\eta}$ as a function of $(J-J_c)L^{1/\nu}$ in the relativistic limit and as a function of $(\lambda - \lambda_c) L^{1/\nu}$  for the Hamiltonian limit in $d=2,3$ dimensions. }
\label{fig:Z2-critical-scaling}
\end{figure*}

\section{Results}
\label{sec:results}

We study our qubit model in both the Relativistic and Hamiltonian limits in $d=2$ and $d=3$ to show that we reproduce the results expected from traditional models near the quantum critical points.  We measure the current-current susceptibility $\rho_s$ and the susceptibility of the two-point correlation function $\chi$, defined in \cref{sec:worm-algorithm}.  For the relativistic limit, we tune the coupling $J$ close to the critical value $J_c$.  Near this quantum critical point, in scaling regime we expect the observables to behave as
\begin{align}
\rho_s L^{d-1} & = f( (J-J_c) L^{1/\nu}),\\
\chi/L^{2-\eta} & =  g((J-J_c) L^{1/\nu}),
\label{eq:critical-scaling}
\end{align}
where $f(x)$ and $g(x)$ are universal functions, and $\nu$ and $\eta$ are the critical exponents. Here we neglect corrections to scaling for simplicity and make sure our data fits to this expected form.   We can extract the critical exponents $\nu, \eta$ and the critical coupling $J_c$ by approximating $f(x)$ and $g(x)$ as fourth order polynomials (including $x^4$) and performing a simultaneous fit of $\rho_s$ and $\chi$ to the above expression, allowing the critical exponents $\nu, \eta$, the critical coupling $J_c$ and the coefficients of polynomial expansion of $f(x)$ and $g(x)$ to vary.  For the Hamiltonian limit, we similarly tune the coupling $\lambda = J_t/J$ (keeping $J$ fixed) to its critical value $\lambda_c$, so we expect
\begin{align}
\rho_s L^{d-1} & = f( (\lambda-\lambda_c) L^{1/\nu}),\\
\chi/L^{2-\eta} & =  g((\lambda-\lambda_c) L^{1/\nu}).
\label{eq:critical-scalingH}
\end{align}
near the critical point.   Even though the functions $f(x)$ and $g(x)$ are universal, unfortunately it is difficult to compare the ones obtained from the relativistic limit to the Hamiltonian limit because the functions also depend on the aspect ratio of the lattice. A complete listing of our results for the fits in various models in $d=2,3$ in both the relativistic and Hamiltonian limits is given in \cref{tab:results}.

\subsection{\texorpdfstring{Results for the $O(3)$ model}{Results for the O(3) model}}

The results for the $O(3)$ model in both $d=2$ and $d=3$ are shown in \cref{fig:O3-critical-scaling}.  In $d=2$, the main result to show is that we can reproduce the physics of the Wilson-Fisher fixed point.  The critical exponents $\nu = 0.7113(11)$, $\eta=0.0378(6)$ for the $O(3)$ are well known critical exponents and have been computed in the literature using the traditional model \cite{Pelissetto:2000ek}.  If we assume that $f(x)$ and $g(x)$ are given by a fourth order polynomials, we find that we can fit our data for both $\rho_s$ and $\chi$ to \cref{eq:critical-scaling} very well, as is shown by the first row of \cref{fig:O3-critical-scaling} for the relativistic limit.

We repeat the above analysis in the Hamiltonian limit by fixing $\varepsilon=0.1$ and varying $\lambda$. In order to mimic cubical boxes we choose $\beta= L$, which means the number of temporal lattice sites now are ten times larger. This makes these computations more time consuming.  We again get excellent fits as shown in the second row of \cref{fig:O3-critical-scaling}.   These results provide strong evidence that the Wilson-Fisher fixed point of the $O(3)$ scalar field theory can be obtained very well using our four qubit quantum Hamiltonian.

In $d=3$ dimensions, the $O(3)$ scalar field theory is free up to logarithmic corrections.  
We thus expect our model to reproduce the mean-field critical exponents.  We show that the results are indeed consistent with the mean-field predictions, as shown in the bottom two rows of \cref{fig:O3-critical-scaling} for the relativistic and Hamiltonian limits.

\subsection{Extension to \texorpdfstring{$O(2)$}{O2} and
\texorpdfstring{$Z_2$}{Z2} models}
\label{sec:extension-O2-Z2}

As described in earlier in \cref{sec:qubit-model}, we can modify the $O(3)$ model to get $O(2)$ and $Z_2$ models as well.  This means we should be able to recover the XY critical exponents in $d=2$ for the $O(2)$ model and the Ising critical exponents in $d=2$ for the $Z_2$ model.

The critical exponents for the $XY$ universality are known from Monte Carlo studies to be $\nu=0.6717(1)$ and $\eta=0.0381(2)$ \cite{campostrini_critical_2006}. \Cref{fig:O2-critical-scaling} shows our results for the $O(2)$ qubit model in $d=2$ spatial dimensions, for the relativistic limit and the Hamiltonian limit (with $\varepsilon=0.1$).  The extracted critical exponents are in good agreement with the literature.  On the other hand, the bottom two rows show the results in $d=3$, where we find the critical exponents to be consistent with the mean-field predictions of $\nu=0.5$ and $\eta=0.0$.

For the $Z_2$ model, we show our results in \cref{fig:Z2-critical-scaling}.  Since there is no analog of $\rho_s$ in this case, we only show the plots for the susceptibility $\chi$ (defined for $Z_2$ with $m=0$).  After tuning the coupling close to the critical point, we perform a single fit of $\chi$ to the form
\begin{align}
\chi & = L^{2-\eta} g(x)
\label{eq:critical-scaling-Z2}
\end{align}
where $x=(J-J_c)L^{1/\nu}$ in the relativistic limit and  $x=(\lambda - \lambda_c) L^{1/\nu}$ for the Hamiltonian limit.   The most precise estimates for the Ising critical exponents in $d=2$ come from conformal bootstrap \cite{Kos:2016ysd}, which gives $\nu = \num{0.629971(4)}$ and $\eta = \num{0.036298(2)}$ in $d=2$.  Once again, we find our results to be consistent with these in both the relativistic and the Hamiltonian limits, as shown in the top row of \cref{fig:Z2-critical-scaling}.   The bottom row of that figure shows our results for $d=3$, which are consistent with the mean-field predictions.

\subsection{Monomer density}
\label{sec:monomer-density}

We have also measured the monomer density $v$, as described in \cref{sec:worm-algorithm}. Its value at the critical point, extracted from an extrapolation of our data on the largest lattice sizes, are shown in the last column of \cref{tab:results}. In the limit $\beta\rightarrow \infty$ the monomer density is the probability of a space-time site to be in the trivial Fock vacuum state. If we write the ground state of our qubit model as
\begin{align}
|\Psi\rangle = |{\bf r},s\rangle \otimes |\Phi^s_{\rm universe}\rangle + \sum_{m} |{\bf r},m\rangle \otimes |\Phi^m_{\rm universe}\rangle,
\end{align}
where $|\Phi^s_{\rm universe}\rangle$ and $|\Phi^m_{\rm universe}\rangle$ are the kets of the universe without the site $\bf r$, then
\begin{align}
v =   \langle \Phi^s_{\rm universe}
|\Phi^s_{\rm universe}\rangle = 
\langle \Psi |{\bf r},s\rangle\langle {\bf r}, s|\Psi\rangle.
\end{align}
We see that it is a measure of how perturbative the ground state is. When $v$ is close to $1$ the theory becomes more and more perturbative. Since the $d=3$ qubit models are described by free field theories close to the critical point, the observed values for the monomer density are seen to be much closer to $1$ than $d=2$ which is known to be less perturbative.

\begin{table*}[!t]
  \SetTableProperties{}
  \setlength{\tabcolsep}{4pt}
  \input{table-results-all.tex}
  \caption{Results for all models.  The relativistic and Hamiltonian limits ($\varepsilon=0.1$) are indicated for each model by (R) and (H), respectively.  For comparison, we include existing results from Monte Carlo computations for $O(3)$ and $O(2)$ in $d=2$ \cite{Pelissetto:2000ek}, conformal bootstrap for $Z_2$ in $d=2$ \cite{Kos:2016ysd} , and mean-field theory for all cases in $d=3$.  }
  \label{tab:results}
\end{table*}

\section{Conclusions}
\label{sec:conclusions}

In this work, we have defined the concept of qubit regularization of quantum field theories and have argued that this is an important step in studying quantum field theories on a quantum computer. Using the example of the $O(3)$ sigma model in $d=2$ and $d=3$, we have argued that qubit regularized models can reproduce the same physics as the traditional lattice regularized quantum field theories in the continuum limits. In particular, we showed that the scaling of the Wilson-Fisher fixed points and the Gaussian fixed points are reproduced accurately from the qubit regularized model involving only two qubits on each lattice site.  We were also able to demonstrate that qubit regularizations constructed for $Z_2$ and $O(2)$ quantum field theories, by suitably modifying the $O(3)$ model, can also reproduce the correct critical exponents.

In our work, we did not consider if our qubit regularization reproduces the physics of the traditional lattice regularized model in \cref{eq:tradmodel} for $d=1$. This problem has been studied recently using the tensor networks within a different qubit regularization scheme \cite{PhysRevD.99.074501}. Based on those results one might conclude that our two qubit model will not lead to a viable qubit regularization. However, we believe the case of $1+1$ dimensions is more subtle. While the traditional model has a critical point at $g_c=0$ where the asymptotically free $O(3)$ quantum field theory emerges, it may be naive to expect that something similar would happen with all qubit regularizations. Since the $O(3)$ symmetry would prevent the superfluid phase to form due to the Mermin-Wagner theorem, there is of course a possibility that there is no critical point in the theory constructed with two qubits, as found in \cite{PhysRevD.99.074501}. On the other hand in our qubit model, we can guarantee the existence of a critical point at least at $\lambda_c=-\infty$ like in the traditional model. This is due to the emergence of a new $U(1)$ symmetry present in all loop models on bi-partite lattices. Previous work shows that our loop model will be in the critical Kosterlitz-Thouless phase. \cite{PhysRevD.77.054502}. So our qubit model seems to be different from what was studied earlier. In fact, the issue that makes things more subtle for us is that we cannot rule out a topological phase transition to an $O(3)$ symmetric Kosterlitz-Thouless phase even at a finite value of $\lambda_c$. If this occurs, there could be new fixed point that emerges at $\lambda_c < -\infty$ in our qubit regularized model. At this critical point we may either obtain the usual asymptotically free $O(3)$ quantum field theory or something more exotic. To sort this out, we postpone the $d=1$ study to a future publication.

\section*{Acknowledgments}

We thank Ribhu Kaul and Uwe-Jens Wiese for helpful conversations. We also thank Tanmoy Bhattacharya, Rajan Gupta and Rolando Somma for helpful discussions and critical feedback during the project. The material presented here is funded in part by a Duke subcontract from Department of Energy (DOE) Office of Science - High Energy Physics Contract \#89233218CNA000001 to Los Alamos National Laboratory and by U.S. Department of Energy, Office of Science, Nuclear Physics program under award Number DE-FG02-05ER41368. 

\appendix

\section{Tests of the Algorithm}

We test our Monte Carlo algorithm in various ways. First, we compute the four observables discussed in the text exactly starting from the definition of the lattice partition function \cref{eq:latpf} on a $2 \times 2$ lattice by enumerating all possible configurations. For the $O(3)$ model, the partition function is given by
\begin{align}
Z \ & = \ 1+ 6 W_s^2 + 2 W_t^2 (1+W_\mu^2 + W_\mu^{-2}) + 9 W_s^4
\nonumber \\
& + W_t^4 (1+W_\mu^2 + W_\mu^{-2})^2 
+ 8 W_s^2 W_t^2 (4 + W^2_\mu + W_\mu^{-2}), 
\end{align}
where the weights $W_t$, $W_s$ and $W_\mu$ are defined in \cref{eq:weights}. 
Since this is the $O(3)$ model, the above expression includes terms from both oriented and unoriented loops.  We can write down a similar expression for the $O(2)$ or $Z_2$ model by omitting unoriented or oriented loops, respectively.  
In \cref{tab:exact-2x2}, we show a comparison between the exact results and those obtained using our Monte Carlo method for all three ($O(3)$, $O(2)$ and $Z_2$) models.

\begin{table*}[htb]
\SetTableProperties{}
\sisetup{input-protect-tokens=\dots}
\input{table-exact2x2.tex}
\caption{Comparison of exact results and the results obtained from Monte Carlo methods on a $2 \times 2$ space-time lattice for the $O(3)$, $O(2)$ models. In the presence of a chemical potential, $\chi$ as defined in \cref{eq:chidef} depends on $m$:  $m=1$ for $O(3)$ and $O(2)$, and $m=0$ for $Z_2$.
\label{tab:exact-2x2} }
\end{table*}

\begin{figure*}[htb]
\centering
\includegraphics[width=0.9\linewidth]{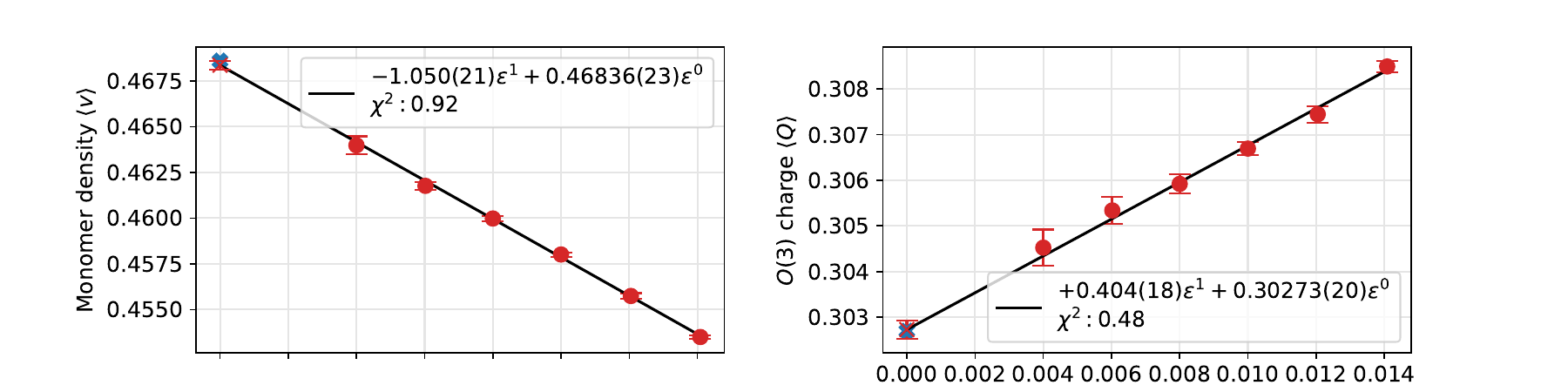}
\caption{Continuum time extrapolation of monomer density $v$ and the $O(3)$ charge $\langle Q\rangle$ for the $d=1$ data shown in the middle row ($L=4$) of \cref{tab:exact-hamiltonian}. }
\label{fig:epsilon-extrapolation}
\end{figure*}

We also check that our formulation reproduces the qubit Hamiltonian in the $\varepsilon J \rightarrow 0$ limit by explicitly diagonalizing the Hamiltonian and computing the $O(3)$ charge $\langle Q\rangle$ and the monomer density $\< v \> $ in one spatial dimensions on small lattices.  For this, we perform calculations at several finite values of $\varepsilon$ and then perform a linear extrapolation to $\varepsilon\to0$. In \cref{tab:exact-hamiltonian}, we show results from this procedure compared with the exact results from an explicit diagonalization.   We show an illustrative fit to the continuous time limit $\varepsilon\to0$ in \cref{fig:epsilon-extrapolation}.

\begin{table*}[htb]
\sisetup{input-protect-tokens=\dots}
\SetTableProperties{}
\setlength{\tabcolsep}{12pt}
\input{table-exact-hamiltonian.tex}
\caption{Results from diagonalizing the Hamiltonian \cref{eq:latticemodel}.\label{tab:exact-hamiltonian}}
\end{table*}

\bibliography{ref}

\end{document}

%% file: table-results-all.tex
\begin{tabular}{c c S[table-format=2.10]@{\hspace{1em}} @{\hspace{1em}}S[table-format=2.10]  *{2}{S[table-format=3.7]} S[table-format=1.2] @{\hspace{0.0em}} @{\hspace{3em}} S[table-format=2.7] S[table-format=2.7]}
\TopRule
&  & & \multicolumn{4}{c}{Fit} & \multicolumn{2}{c}{Literature} \\
Model & $d$ & \multicolumn{1}{c}{$\< v \>\quad $} & \multicolumn{1}{c}{$J_c$ or $\lambda_c$} & \multicolumn{1}{c}{$\nu$} & \multicolumn{1}{c}{$\eta$} & \multicolumn{1}{c}{$\chi^2/\text{d.o.f}\quad\quad$} & $\nu$ & $\eta$\\
\HeaderRule
$O(3)$ (H) & 2 & 0.785956(18) & 4.81695(37) & 0.693(15) & 0.038(26) & 0.32 & 0.7113(11) & 0.0378(6)\\
$O(3)$ (R) & 2 & 0.78162325(68) & 0.244327(10) & 0.7048(70) & 0.031(12) & 0.52 & 0.7113(11) & 0.0378(6)\\
$O(3)$ (H) & 3 & 0.9094594(89) & 10.09751(61) & 0.497(12) & -0.043(60) & 0.41 & 0.5 & 0.0\\
$O(3)$ (R) & 3 & 0.920674(11) & 0.1558650(31) & 0.4945(64) & -0.004(31) & 0.34 & 0.5 & 0.0\\
\MidRule
$O(2)$ (H) & 2 & 0.854411(21) & 5.462676(88) & 0.6707(37) & 0.0276(86) & 0.34 & 0.6717(1) & 0.0381(2)\\
$O(2)$ (R) & 2 & 0.852320(16) & 0.2320373(36) & 0.6626(38) & 0.0308(71) & 0.89 & 0.6717(1) & 0.0381(2)\\
$O(2)$ (H) & 3 & 0.9396982(66) & 10.55476(66) & 0.516(11) & 0.091(63) & 0.52 & 0.5 & 0.0\\
$O(2)$ (R) & 3 & 0.946779(25) & 0.1529526(21) & 0.4921(43) & -0.028(23) & 0.52 & 0.5 & 0.0\\
\MidRule
$Z_2$ (H) & 2 & 0.925073(33) & 6.05974(21) & 0.6416(95) & -0.016(45) & 0.48 & 0.629971(4) & 0.036298(2)\\
$Z_2$ (R) & 2 & 0.924076(60) & 0.2219655(45) & 0.6375(50) & -0.037(18) & 1.69 & 0.629971(4) & 0.036298(2)\\
$Z_2$ (H) & 3 & 0.98362920(87) & 10.9929(30) & 0.507(26) & -0.28(38) & 0.39 & 0.5 & 0.0\\
$Z_2$ (R) & 3 & 0.973115(33) & 0.150208(18) & 0.471(14) & -0.02(26) & 1.22 & 0.5 & 0.0\\
\MidRule
\BotRule
\end{tabular}

%% file: table-exact2x2.tex
\begin{tabular}{ *{9}{c} }
\TopRule
Model & $d$ & $\varepsilon J$ & $\lambda$ & $\mu/J$ & $v$ & $\chi$ & $Q$ & $\rho_s$  \\
\HeaderRule

\multirow{8}{*}{$O(3)$} & 1 & \tablenum{0.1} & \tablenum{0.1} & \tablenum{0.5} & \tablenum{0.24656(20)} & \tablenum{0.84704(22)} & \tablenum{0.12285(16)} & \tablenum{0.09704(14)} \\
  &   &   &   &   & \tablenum{0.246650\dots} & \tablenum{0.847499\dots} & \tablenum{0.122955\dots} & \tablenum{0.097111\dots} \\

  & 1 & \tablenum{0.1} & \tablenum{10} & \tablenum{5} & \tablenum{0.62100(19)} & \tablenum{1.40391(28)} & \tablenum{0.26977(39)} & \tablenum{0.39735(24)} \\
  &   &   &   &   & \tablenum{0.620923\dots} & \tablenum{1.404305\dots} & \tablenum{0.269603\dots} & \tablenum{0.397517\dots} \\

  & 1 & \tablenum{0.01} & \tablenum{100} & \tablenum{150} & \tablenum{0.25890(22)} & \tablenum{0.71727(17)} & \tablenum{0.01078(16)} & \tablenum{1.40480(24)} \\
  &   &   &   &   & \tablenum{0.259008\dots} & \tablenum{0.717318\dots} & \tablenum{0.010723\dots} & \tablenum{1.404658\dots} \\

  & 1 & \tablenum{0.5} & \tablenum{2.1} & \tablenum{1.8} & \tablenum{0.32459(12)} & \tablenum{1.26793(39)} & \tablenum{0.67913(27)} & \tablenum{0.51253(19)} \\
  &   &   &   &   & \tablenum{0.324648\dots} & \tablenum{1.268206\dots} & \tablenum{0.679488\dots} & \tablenum{0.512282\dots} \\
\MidRule

\multirow{8}{*}{$O(2)$} & 1 & \tablenum{0.1} & \tablenum{0.1} & \tablenum{0.5} & \tablenum{0.32586(19)} & \tablenum{1.17026(62)} & \tablenum{0.21541(50)} & \tablenum{0.13045(84)} \\
  &   &   &   &   & \tablenum{0.325838\dots} & \tablenum{1.170464\dots} & \tablenum{0.215784\dots} & \tablenum{0.128823\dots} \\

  & 1 & \tablenum{0.1} & \tablenum{10} & \tablenum{5} & \tablenum{0.68294(30)} & \tablenum{1.57691(78)} & \tablenum{0.32413(86)} & \tablenum{0.44085(34)} \\
  &   &   &   &   & \tablenum{0.682897\dots} & \tablenum{1.577283\dots} & \tablenum{0.324587\dots} & \tablenum{0.440493\dots} \\

  & 1 & \tablenum{0.01} & \tablenum{100} & \tablenum{150} & \tablenum{0.26872(39)} & \tablenum{0.74361(34)} & \tablenum{0.01171(23)} & \tablenum{1.45513(79)} \\
  &   &   &   &   & \tablenum{0.268415\dots} & \tablenum{0.743762\dots} & \tablenum{0.011517\dots} & \tablenum{1.455718\dots} \\

  & 1 & \tablenum{0.5} & \tablenum{2.1} & \tablenum{1.8} & \tablenum{0.35538(19)} & \tablenum{1.49944(51)} & \tablenum{0.78810(52)} & \tablenum{0.62482(31)} \\
  &   &   &   &   & \tablenum{0.355329\dots} & \tablenum{1.499587\dots} & \tablenum{0.787636\dots} & \tablenum{0.625324\dots} \\
\MidRule

\multirow{8}{*}{$Z_2$} & 1 & \tablenum{0.1} & \tablenum{0.1} & -- & \tablenum{0.48536(17)} & \tablenum{1.8955(11)} & -- & -- \\
  &   &   &   &   & \tablenum{0.485637\dots} & \tablenum{1.895648\dots} & -- & -- \\

  & 1 & \tablenum{0.1} & \tablenum{10} & -- & \tablenum{0.860500(89)} & \tablenum{1.96583(78)} & -- & -- \\
  &   &   &   &   & \tablenum{0.860676\dots} & \tablenum{1.964825\dots} & -- & -- \\

  & 1 & \tablenum{0.01} & \tablenum{100} & -- & \tablenum{0.880497(81)} & \tablenum{1.57524(87)} & -- & -- \\
  &   &   &   &   & \tablenum{0.880590\dots} & \tablenum{1.575397\dots} & -- & -- \\

  & 1 & \tablenum{0.5} & \tablenum{2.1} & -- & \tablenum{0.59356(33)} & \tablenum{2.5946(14)} & -- & -- \\
  &   &   &   &   & \tablenum{0.593563\dots} & \tablenum{2.593807\dots} & -- & -- \\
\MidRule
\BotRule
\end{tabular}

%% file: table-exact-hamiltonian.tex
\begin{tabular}{*{5}{c} *{2}{S[table-format=2.10]}}
\TopRule
 \multicolumn{1}{c}{$d$}  &  \multicolumn{1}{c}{$L$}  &  \multicolumn{1}{c}{$\lambda$}  &  \multicolumn{1}{c}{$\mu/J$}  &  \multicolumn{1}{c}{$J \beta$}  &  \multicolumn{1}{c}{$v$}  &  \multicolumn{1}{c}{$Q$} \\
\HeaderRule
1 & 2 & 0.50 & 0.30 & 1.0 & 0.49608(15) & 0.11168(10)\\
  &   &   &   &   & 0.496259\dots & 0.111483\dots\\
1 & 4 & 0.50 & 0.30 & 1.0 & 0.46836(23) & 0.30273(20)\\
  &   &   &   &   & 0.468597\dots & 0.302696\dots\\
1 & 6 & 0.50 & 0.30 & 1.0 & 0.45990(25) & 0.47591(22)\\
  &   &   &   &   & 0.460300\dots & 0.475203\dots\\
\BotRule
\end{tabular}